# Intrinsic rotation reversal, non-local transport, and turbulence transition in KSTAR L-mode plasmas

Y.J.Shi[1], J.M. Kwon[2], P.H.Diamond[3], W.H.Ko[2], M.J.Choi[2], S.H.Ko[2], S.H.Hahn[2], D.H.Na[1], J.E.Leem[4], J.A.Lee[4], S.M.Yang[1], K.D.Lee[2], M.Joung[2], J.H.Jeong[2], J.W.Yoo[2], W.C.Lee[2], J.H.Lee[2], Y.S.Bae[2], S.G.Lee[2], S.W.Yoon[2], K. Ida[5], and Y.S.Na[1]

[1]Department of Nuclear Engineering, Seoul National University, Seoul, Korea
[2]National Fusion Research Institute, Daejeon, Korea
[3]CMTFO and CASS, University of California, San Diego, USA
[4]Pohang University of Science and Technology, Pohang, Korea
[5]National Institute of Fusion Science, Toki, Japan

*E-mail contact of main author:* yjshi@snu.ac.kr     yjshi@ipp.ac.cn

**Abstract**. Experiments of electron cyclotron resonance heating (ECH) power scan in KSTAR tokamak clearly demonstrate that both the cut-off density for non-local heat transport (NLT) and the threshold density for intrinsic rotation reversal can be determined by the collisionality. We demonstrate that NLT can be affected by ECH, and the intrinsic rotation direction follows the changes of NLT. The cut-off density of NLT and threshold density for rotation reversal can be significantly extended by ECH. The poloidal flow of turbulence in core plasma is in the electron and the ion diamagnetic direction in ECH plasmas and high density OH plasma, respectively. The auto-power spectra of density fluctuation are almost the same in the outer region for both ECH and OH plasmas. On the other hand, in the core region of ECH plasmas, the power spectra of the density fluctuations are broader than those of OH plasma. All these observations in macroscopic parameters and micro fluctuations suggest a possible link between the macro phenomena and the structural changes in micro-fluctuations.

## 1. Introduction

There are several outstanding experimental mysteries in magnetic fusion plasma research. A rapid increase of the central electron temperature caused by the edge cooling in Ohmic plasmas, the so called non-local transport (NLT) [1] is one of these challenging issues. Although NLT was observed in the TEXT tokamak 21 years ago, and in many fusion devices afterwards [2-11], the underlying physics mechanism is still unclear. One typical characteristic of NLT is this effect disappears with increasing electron density. The critical density where the NLT effect disappears is defined as cutoff density (CUD). On the other hand, the reversal or bifurcation of the intrinsic plasma rotation found recently [12-16] is another mystery in tokamak plasma transport.   Similar to NLT events, the intrinsic rotation reversal in ohmic heating (OH) plasma is also related to a threshold density. The toroidal rotation at the center of the plasma reverses its direction when the plasma density exceeds certain level in density ramping-up OH plasma experiments [13-16]. Intrinsic rotation is particularly important for International Thermonuclear Experimental Reactor (ITER), since neutral beam injection (NBI) is not sufficient to drive the requisite rotation on ITER and future reactors. It is still challenging to understand heat transport fully and origin of intrinsic rotation in magnetic fusion plasma, which is very important for predictions of the performance of burning plasma tokamaks like ITER.

In the latest KSTAR experiment, we found that the cut-off density of NLT can be significantly extended by applying ECH. At the same time, the threshold density of intrinsic rotation reversal is also increased with ECH. The increment of cut-off density for NLT or intrinsic rotation reversal is proportional to the injecting power of ECH. KSTAR's ECH experiments in this paper show the close correlation between NLT and intrinsic rotation reversal. We try to elucidate possible physics mechanism for NLT and rotational reversal from both the experimental fluctuation results and gyro-kinetic stability analysis.

This paper is organized as follows: first, an introduction to the experiment setup in KSTAR is given in section 2, which includes heating schemes and diagnostics, and relevant plasma parameters. The main experimental results are described in section 3. Discussions on possible physical mechanisms for the intrinsic rotation behavior and non-local transport are presented in section 4. Gyro-kinetic analysis of micro-stability is given in section 5.

## 2. Experimental setup

All the results presented in this paper were obtained in KSTAR superconducting tokamak [17]. Charge exchange recombination spectroscopy (CES) [18] is one of main diagnostics for the measurement of ion temperature (Ti) and toroidal rotation velocity (Vϕ). The system has 32 toroidal lines of sight and can provide complete plasma profiles from the plasma edge to the magnetic axis. The x-ray imagingcrystal spectrometer (XICS) [19] in KSTAR also can measure Ti and Vϕ. Unfortunately, XICS in KSTAR provided one channel of data for the core plasma during the last three year's experiments. The profile data for Ti and Vφ in this paper were obtained using CES. Electron temperature was measured with the electron cyclotron emission radiometer (ECE) [20]. The line averaged density measurements are based on microwave interferometry. Thomson scattering [21] provided localized density information for some shots in this paper. Moreover, the electron cyclotron emission imaging system (ECEI) [22] can provide the information of temperature fluctuation. Microwave imaging reflectometry (MIR) [23] measure the density fluctuation in KSTAR.

Neutral beam injection (NBI) beam blip is used for the measurement of CES. Multi-pulse supersonic molecular beam injection (SMBI) [24] is applied to trigger NLT and find the cut-off density. Simple structure, good flexibility and relatively high fuelling efficiency make SMBI the most suitable tool to investigate NLT phenomenon which has been firstly applied to fusion plasma by Lianghua Yao in HL-1M tokamak [25]. There are three ECH heating systems in KSTAR [26]. One is 110 GHz (source power of 0.5 MW) and the others are 170 GHz (source power of 1 MW) and 140/105 GHz (source power of 1.0 MW with dual frequency). The experimental data in this letter was obtained with the 170 or 140 GHz ECH heating system. The ECH system was configured for an X-mode second harmonic and the deposition location was on-axis position.

The main plasma parameters in this experiment are $I_p$ = 0.5 MA, $B_T$ = 2.9 T for 170 GHz ECH and 2.5 T for 140 GHz ECH. Discharges were operated in elongated limiter configurations. Plasma elongation and the edge safety factor are κ = 1.35-1.45, $q_{95}$ = 6.7-7.1 and κ = 1.61-1.65, $q_{95}$ = 7.2-7.5 for $B_T$ = 2.9 T and $B_T$ = 2.5 T, respectively.

## 3. Experimental results

In 2015 KSTAR experiment, we found that the cut-off density of NLT can be significantly extended by applying 170 GHz ECH at $B_T$ = 2.9T. Fig.1 shows the waveforms of a representative ECH discharge and OH discharge in KSTAR. The two shots in fig.1 have the same plasma operation parameters such as plasma current and magnetic field. The difference only results between the two shots results from the injection of 700kW on-axis ECH at the current flattop phase in one discharge. It is clear that the NLT effect (core $T_e$ rise while edge $T_e$ drops) appears in the first SMBI pulse and gradually weakens in later SMBI pulses with increasing density in ECH plasma. Still, very weak NLT can be observed in the last SMBI pulse at $\bar{n}_e$ = 2.4×10$^{19}$m$^{-3}$ in this ECH plasma. On the other hand, NLT cannot be triggered by the first SMBI pulse even at relatively low density of $\bar{n}_e$ = 1.5×10$^{19}$m$^{-3}$ in the OH plasma (the blue curves in fig.1). Due to limited machine time, we couldn't measure the exact cut-off density for the OH plasma in 2015 KSTAR campaign.

In addition to the NLT event, we found another macroscopic difference in the ECH and the OH plasmas. Fig.2 compares the equilibrium profiles of the ECH and OH plasmas presented in fig.1 at same density level ($\bar{n}_e = 2.0 \times 10^{19} m^{-3}$.). It is well known that ECH provides localized electron heating. As a result, the increase of electron temperature and its gradient at the center is observed as expected in the on-axis ECH case (see fig.2a). On the other hand, the ion temperature profiles show little change. One notable point in fig.2 is the core toroidal rotation, which is in the counter-current direction in the OH plasma whereas the co-current direction in ECH plasma. As observed in previous experiments, the intrinsic rotation reversal in OH plasma is normally achieved by increasing the electron density [13-16]. However, here we use the ECH to achieve a very similar phenomenon of intrinsic rotation reversal, as reported in the literature [13-16].

Fig. 3 shows the detailed time trace of $T_e$ at different radii for one typical NLT event in the ECH plasma. The response of $T_e$ rise in the core is about 16ms delayed after the Te drop at the edge, which is much shorter than the energy confinement time (~100ms) for this plasma. Although sawtooth instabilities appear in this discharge, the NLT-induced core $T_e$ increase shows a longer time scale than the sawtooth time scale, and the sawtooth activities do not change the overall NLT trends.

We tried to repeat the similar experiment in 2016 KSTAR campaign, but using the new 140GHz ECH system with $B_T$ = 2.5T. The same trend of NLT has been found with this experimental condition but increment of cut-off density for NLT and the threshold density for rotation reversal is newly found to be proportional to the ECH power as shown in fig.4. The CUD is only $1.21 \sim 1.25 \times 10^{19} m^{-3}$ for the OH plasma, but it is dramatically increased to $2.2 \sim 2.3 \times 10^{19} m^{-3}$ by 800 kW ECH in the ECH plasma. On the other hand, the normalized effective collisionality ($\nu_{eff}$) at half normalized radius ($\rho \sim 0.5$) at cut-off density is almost the same for the four OH and ECH shots.

The time evolution of the core toroidal rotation is shown in fig.5. The data in fig.5 and fig.4 were obtained from the same discharges. ECH is injected from 3s to 7s for three ECH shots. For the OH shot 16471, the core $V\phi$ is in the co-current direction in the initial low density phase, and gradually decreases and changes to the counter-current direction as the density increases. For the three ECH shots, there appear very clear increments of the core $V\phi$ in the co-current direction as ECH is injected. However, very interestingly, the core $V\phi$ decreases as the density reaches a certain level in the ECH plasmas. Although the absolute value and the ramp-up rate of density are almost identical for the three ECH shots, the threshold density for the core $V\phi$ reversal (co-direction→ counter-direction) is obviously higher for higher ECH power heating. The modulation frequency of NBI blips is 2Hz. So the exact timing of the core rotation reversal is hard to identify in these plasmas. But we can estimate the threshold density for the rotation reversal from the time evolutions. The estimated threshold density, which is also shown in fig.4, is very similar to the cut-off density for NLT in fig.4. More important finding is that both the cut-off density for NLT and the threshold density for the rotation reversal increase as the ECH power increases. And $v_{eff}$ at rotation reversal point is also quite close the $v_{eff}$ at cut-off point. On the other hand, the maximum value of core $V\phi$ shows little changes for the ECH power variation.

Fig.6 shows the 2-D electron temperature images taken from ECEI during NLT in the ECH plasma shown in in fig.1. Compared to the 1-D ECE signals shown in fig.3, ECEI images can show more intuitional and clearer structures in 2-D space during NLT. The big red spot represents the area with $T_e$ increase during NLT. Here, we call this red spot as the NLT pattern. At the same time, m/n = 1 island, the small blue spot around the q = 1 surface, is also clearly seen inside the NLT pattern, which corresponds to the sawtooth activities identified in the ECE signals (see fig.3). We note that there is no other obvious island structure observed other than this m/n = 1 island. The m/n = 1 island rotates while the whole NLT pattern does not rotate in the time period shown in fig.3. Both the ECEI images in fig.6 and ECE signals in fig.3 indicate that the correlation between the NLT and the sawtooth activity is weak.

The fluctuation characteristics of these ECH and OH plasmas are also investigated by 2-D MIR fluctuation diagnostics. Fig.7 shows the density fluctuation measured with the MIR system. It can be seen that the auto-power spectra of density fluctuations are almost the same for both the ECH and OH plasma in the outer region ($\rho>0.4$). However, in the core region ($\rho<0.3$), the power spectra of the density fluctuation of the ECH plasma are clearly broader than those of the OH plasma. The discrepancy in the density power spectra between the ECH and the OH plasma increases with the frequency, especially at the relative high frequency range.

MIR can also provide the information of poloidal rotation velocity $v_\theta$ [23, 27]. Fig.8 shows the cross-phase spectra of MIR in the core region at a low density ECH phase and a high density OH phase. $v_\theta$ estimated from the dispersion relation of cross-phase spectrum is about 6.5km/s in the electron diamagnetic direction in the ECH phase and 7km/s in the ion diamagnetic direction in the OH phase. According to Ref.24 and 27, the apparent poloidal rotation velocity measured by MIR also includes the contribution from toroidal flow. Here, the effects of $V_\phi$ to $v_\theta$ can be excluded. Firstly, $V_\phi$ in non-NBI heating plasma in this paper is much smaller than that in NBI heating plasma in ref.24 and 27. Secondly, the toroidal flow in the core region is co-current direction in the ECH plasma and counter-current direction in high density OH plasma as shown in previous part. Contribution from toroidal flow is in the ion diamagnetic direction for co-current toroidal rotation and in the electron diamagnetic direction for counter-current toroidal rotation. More, static poloidal rotation velocity estimated with neoclassic calculation [28] is only about 1km/s in the core region for the plasma parameters presented in this paper. So, here, the apparent $v_\theta$ mainly represents the phase velocity of turbulence. Now, the microscopic poloidal rotation reversal of turbulence (electron diamagnetic direction in the ECH plasma, ion diamagnetic in the high density OH plasma) is disclosed for the first time, which is covered by the more obvious macroscopic toroidal flow reversal (co-current in the ECH plasma and counter-current in the high density OH plasma).

## 4. Discussion on physical mechanisms for intrinsic rotation reversal and non-local transport

The experimental observations in the previous section suggest a possible link between the macro phenomena and the structural changes in micro-fluctuations. One prominent theoretical hypothesis to explain the cut-off density for NLT is the transition of dominant fluctuation mode from trapped electron mode (TEM) mode to ion temperature gradient (ITG) related to the confinement regime transition between linear ohmic confinement (LOC) and saturated Ohmic confinement (SOC) [29]. It is known that the collisionality is one key issue to determine the dominated turbulence mode. Fig.4 shows that the normalized effective collisionality at higher density with higher ECH power is the same level as the low density OH plasma. Very interestingly, the intrinsic rotation reversal can also be explained by the transition of the fluctuation or confinement mode [30-31].

Due to the excellent localized heating, ECH is a good control 'knob' for the micro-turbulence population. Although the toroidal rotation driven by ion cyclotron resonance heating (ICRF) and lower hybrid current drive (LHCD) has also been observe in many devices, the mechanism for torque or rotation change induced by ECH is not exactly the same as ICRF or LHCD plasma. ICRF can directly drive toroidal flow with mode conversion [32] or energetic particles accelerated by wave [33]. LHCD-driven rotation is also affected by many actuators which include fast electron ripple loss effects, thermal ripple induced neoclassical friction, and direct LH wave effects [34-35]. ECH almost does not bring such complicated factor as ICRF or LHCD. The extremely low toroidal field ripple in KSTAR (as low as ~ 0.05% [36]) can also further reduce the effect of fast particle ripple loss and the ripple induced neoclassical friction for thermal particle. On the whole, the rotation change by ECH in KSTAR is more likely related to the variation of turbulence induced intrinsic torque.

The relation between the cut-off density for NLT (or threshold density for rotation reversal) and the ECH power strongly implies that the transition or change of micro-turbulence mode is a possible cause for the rotation reversal or the cut-off density for NLT. But there are still several puzzles in the detailed physical processes for the formation of intrinsic toroidal rotation profile. The first is why the core rotation direction flips only, while the edge rotation still in the co-current direction? The second one is why the co-directional core rotation profile in the low density OH plasma has a flatter form than the counter-directional rotation profile in the high density OH plasma. These two characters of the rotation profiles are also observed in fig.2c in this paper. Here, we provide a possible explanation for these puzzles. As shown in fig.9, we consider there are two types of intrinsic torque. The first one is an edge/SOL intrinsic torque, which depends on plasma boundary and SOL conditions. The second torque is located in core plasma and driven by turbulence. The direction of the second torque can be reversed, if the dominant turbulence mode is changed from ITG to TEM or TEM to ITG. We want to emphasize that the direction of ITG/TEM-driven torque is not fixed, and depends on other plasma parameters such as the magnetic shear or plasma current profile [37, 38]. For example, TEM-driven torque can be in counter-direction for plasma with broad current profiles and in co-current direction with monatonic current profiles [37, 38]. For the OH and ECH plasmas with peaked monatonic current profile, ITG/TEM-driven torque can be in counter/co direction. In high density OH plasmas, the combined actions of the counter-direction ITG-driven torque in core plasma and co-direction torque in edge/SOL region can make the reversal happen only in the core.

In low density OH or ECH plasmas, on the other hand, the rotation is in co-direction in the whole region and the rotation profile is nearly flat in the core. There is no doubt that worse confinement in low density OH or ECH plasmas is one contributor for the formation of flat profile. Here, we note that the neoclassical toroidal viscosity (NTV) due to MHD activity [39, 40] can be a damping mechanism. Although, TEM intensity can be enhanced with high electron temperature and its gradient, MHD activities also increase as the electron temperature and its gradient increase. Fig.10 shows the fluctuation power spectrum of Te ($\delta$Te/Te) measured by ECE at $\rho$~0.5. The fluctuation of Te in low frequency domain can represent the intensity of MHD activity. It can been seen that the fluctuations of Te for low density OH plasma (green line in fig.9) is stronger than that of high density OH plasma (blue line in fig.10), which implies that the MHD-induced damping can be bigger in the low density OH plasma and make the core rotation profile more flat. Fig.10 also implies that the MHD damping can increase with the power of ECH. This may explain why the maximum co-rotation value does not increase with the ECH power. Two experimental evidences to support TEM-driven co-direction torque are shown in fig.11 and 12. Fig.11 shows the evolutions of the rotation profile in during a discharge. For the low density ECH plasma (the red line in fig.11), the core rotation reaches a high level. Moreover, the clear gradient of $V\phi$ (peaked profile) appears in the core region in the low density ECH plasma. Fig.12 shows the continuous time evolution of the core $V\phi$ measured by XICS. The parameters for the shot 16478 in fig.12 are almost identical with the shot 16475 in fig.11 except NBI beam blips. It can be seen that there is a prompt change of the core $V\phi$ after the ECH injection. Then there is a slowly increasing phase of the $V\phi$. The prompt change of the core $V\phi$, which is corresponded to the fast drop phase of $\nu_{eff}$ after ECH injection, is much faster than the typical momentum confinement time of about several hundred milliseconds in KSTAR and should be be induced by the enhanced co-direction TEM-driven torque by ECH. The profiles of edge electron temperatures in this shot are shown in fig.13. Both the amplitude and gradient of edge Te increase quite a lot after ECH injection, which may provide higher edge torque for ECH plasma. The slower change can be caused by the enhanced edge/SOL torque through transport mechanisms such as the momentum pinch, which is very similar to the phenomenon in Ref [41].

## 5. Gyro-kinetic linear stability analysis

To investigate the possible change of the turbulence characteristics, numerical studies have been performed for the OH and ECH plasmas using the gyro-kinetic code GYRO [42, 43]. Linear gyro-kinetic simulations have been conducted at the radii from $\rho= 0.25$ to $\rho= 0.55$. These results are presented in fig.12 where the positive and the negative mode frequencies imply the modes propagate in the electron and the ion diamagnetic drift direction, respectively. It can be seen that the TEM is dominant for the whole simulated region in the ECH plasma.   At the outer region ($\rho= 0.55$) both the mode frequency and the growth rate are very similar between the OH plasma and the ECH plasma. On the other hand, the growth rate and the mode frequency of OH plasma become lower than those of the ECH plasma as the radial location moves to the center. Especially, at the deep core region ($\rho= 0.25$), there is no linearly excited mode in the OH plasma while strong TEM still exist in the ECH plasma. The trend of the difference in growth rate between the OH and the ECH plasma with radius is consistent with the experimental observation by MIR in fig.7.

## 6. Summary and future plans

In summary, the correlation between non-locality transport events and rotation reversals has been found in KSTAR L-mode plasmas. The cut-off density for non-locality transport can be increased by ECH heating. The toroidal rotation for the core plasma is reversed from the counter-current direction in the OH plasma to the co-current direction in the ECH plasma. Both the cut-off density for NLT and the threshold density for the core rotation reversal increase with the power of ECH. ECEI image indicates that the NLT phenomenon has weak relation to sawtooth activity. The KSTAR experimental results seem to suggest that the intrinsic toroidal rotation profiles in OH/ECH plasmas are determined by the interplay of (i) an edge/SOL intrinsic torque, which is in co-current direction, (ii) a core intrinsic torque, which is in counter-current direction for high density SOC plasma and co-direction for low density LOC and ECH plasmas, (iii) MHD-induced damping. The microscopic turbulenced poloidal rotation reversal (electron diamagnetic direction in ECH plasma, ion diamagnetic in high density OH plasma) is the powerful experimental evidence for the turbulence mode transition (ITG↔TEM). The linear instability study also shows that strong TEM instabilities are excited in the deep core region of the ECH plasmas. The increase of the electron temperature by ECH can reduce the collisionality for a given density, which impacts the destabilization of TEM. Also, the steepening of Te gradient can help the excitation of TEM. The difference of turbulence characteristics from numerical gyrokinetic simulation of the OH plasma and the ECH plasma also support the fluctuation measurement results with MIR. Although the gyro-kinetic analysis shows the expected trend for linear micro-instabilities, nonlinear gyro-kinetic simulations are necessary to address more quantitative aspects of turbulence change and its effects on intrinsic torque and global profiles. In future works, we also plan to perform nonlinear gyro-kinetic simulations to investigate those subjects and related physical issues.

**Figure Captions**

**Fig.1** The waveforms of two identical ECH and OH shots. (a) line-averaged density, (b) core electron temperature, (c) power of ECH and NBI

**Fig.2** Profiles of macro parameters for two ECH and OH discharges in fig.1 at same density. (a) electron temperature, (b) ion temperature, (c) toroidal rotation velocity,

**Fig.3** Temporal evolutions of electron temperature measured at different radii in the ECH plasma for the first SMBI pulse in fig.1.

**Fig.4** The cut-off density for NLT and threshold density for rotation reversal in OH and ECH power scan shots. The normalized effective collisionality ($\nu_{eff}$) at $\rho \sim 0.5$ at cut-off density and threshold density is also plotted in this figure.

**Fig.5** Temporal evolutions of core $V\phi$ (measured by CES) and line averaged density for OH plasma and ECH plasmas with power scan.

**Fig.6** 2-D electron temperature fluctuation image from ECEI during NLT in the ECH plasma. The time interval for the two image is 0.26 ms. The pattern of temperature increasing area (red colour) is very clearly seen. The black circle represent the approximate location of the q = 1 surface. The blue spot at the q = 1 surface corresponds to the m/n=1 island, which changes its location in time due to the rotation of the island, a general property of the MHD island.

**Fig.7** Power spectra of density fluctuations in the ECH and OH plasmas. The time period in this figure is same as fig.2.

**Fig.8** The cross-phase spectra of MIR. Here, 5 dispersions for centre channels with respect to channel 9 are overlapped. The poloidal velocity is estimated from the inverse of the slope of a fitted line (green colour) on the dispersion relation. The measurement position is about $\rho \sim 0.2$ for the ECH plasma and $\rho \sim 0.35$ for the OH plasma..

**Fig.9** Cartoons shows the torque and damping source for intrinsic rotation in KSTAR

**Fig.10** Fluctuation power spectrum of Te at $\rho \sim 0.5$. The data of 400kW ECH is obtained from shot 16473. The other data are taken from shot 16475.

**Fig.11** Rotation profiles in #16475.

**Fig.12** Temporal evolution of core $V\phi$, density, ECH power, and $\nu_{eff}$ at $\rho \sim 0.5$ in shot 16478.

**Fig.13** Profiles of electron temperature at edge region measured by ECE in shot 16478. ECH is injected at 3.0s. The last closed flux surface is at about R=2.24m estimated by EFIT.

**Fig.14** The mode frequency and the growth rate for the OH plasma and the ECH plasma. The experimental parameters for simulation are taken from fig.3. Positive mode frequency means the modes in the electron diamagnetic direction and negative mode frequency means the modes in the ion diamagnetic direction.

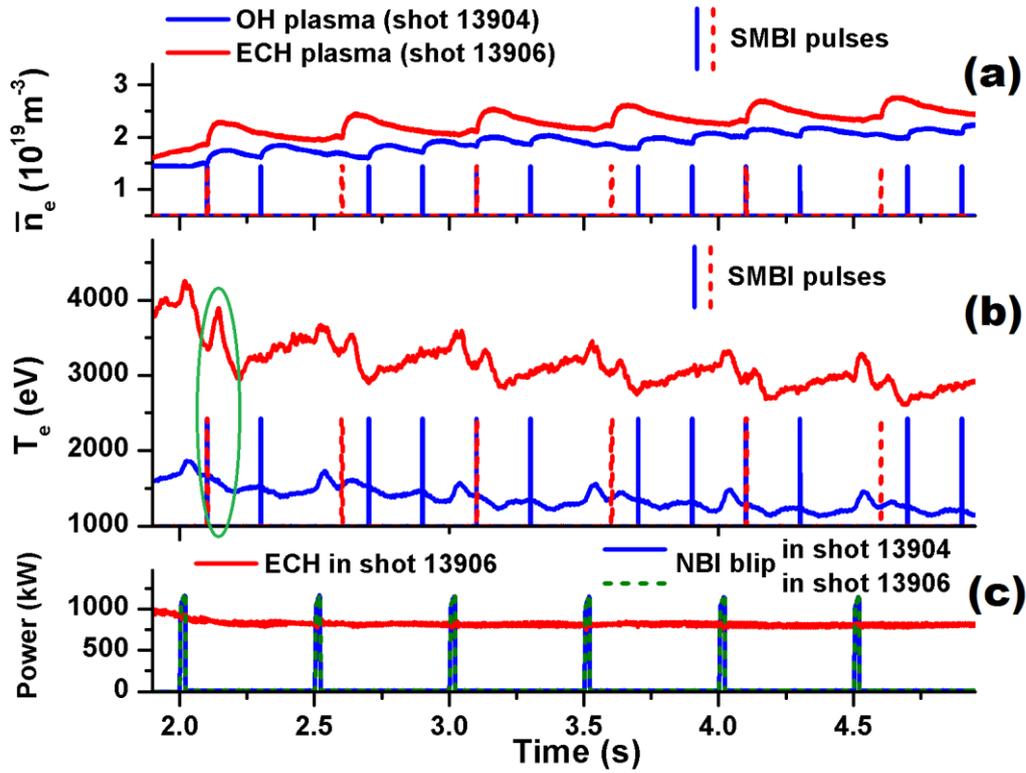

**Fig.1** The waveforms of two identical ECH and OH shots. (a) line averaged density, (b) core electron temperature, (c) power of ECH and NBI

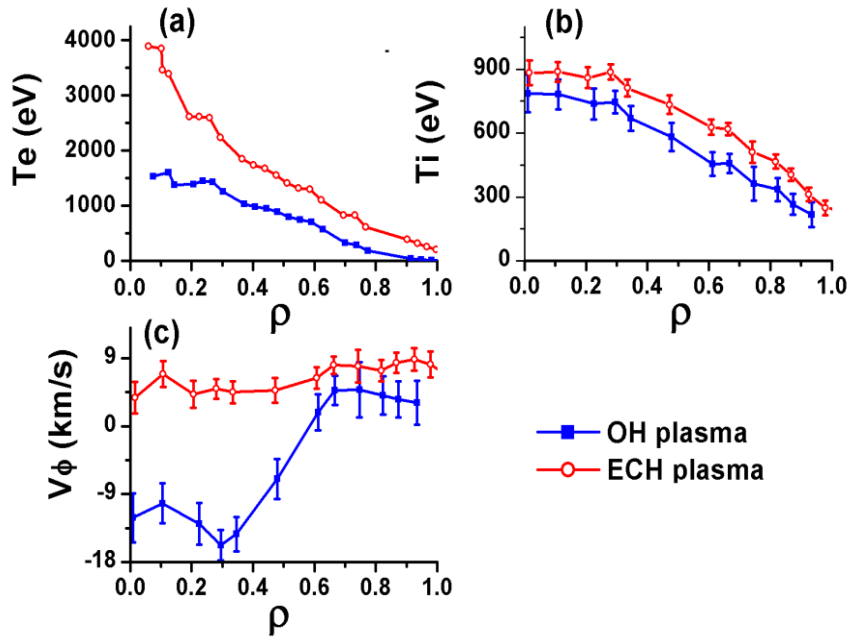

**Fig.2** Profiles of macro parameters for two ECH and OH discharges in fig.1 at same density. (a) electron temperature, (b) ion temperature, (c) toroidal rotation velocity,

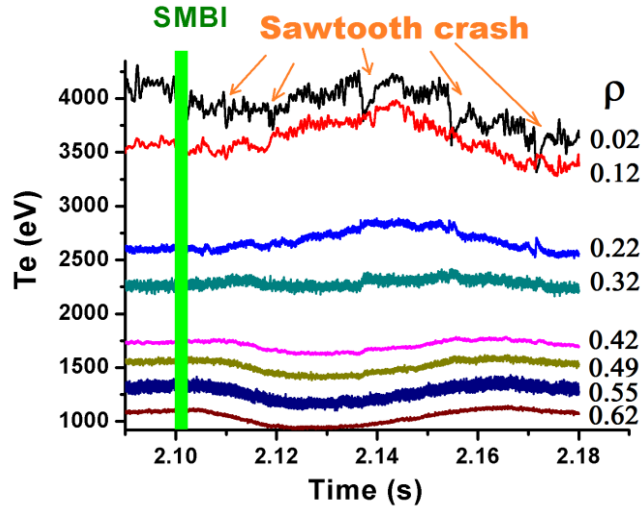

**Fig.3** Temporal evolutions of electron temperature measured at different radii in the ECH plasma for the first SMBI pulse in fig.1.

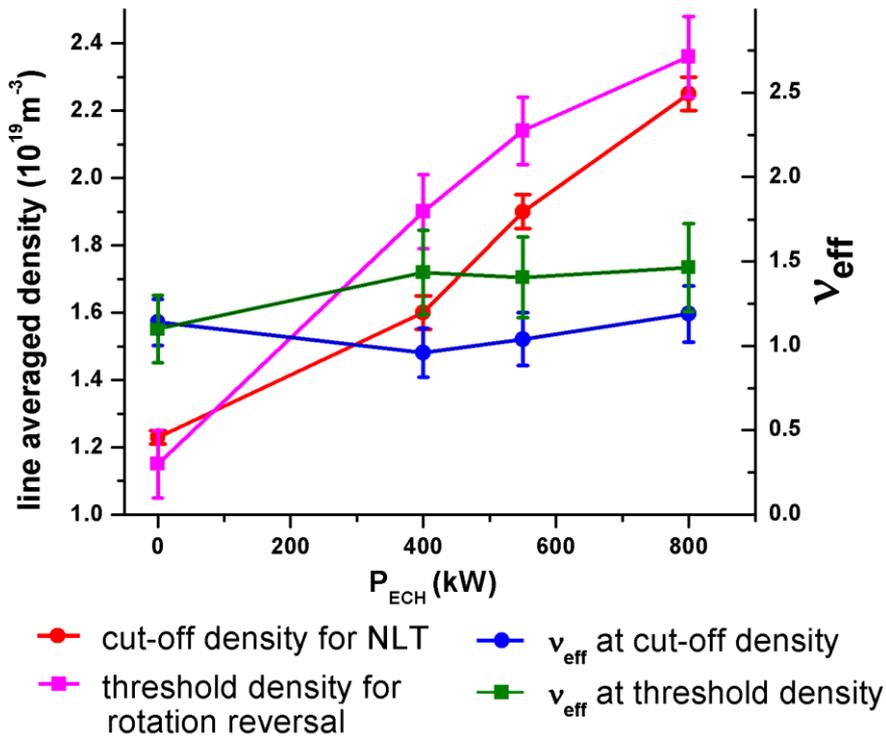

**Fig.4** The cut-off density for NLT and threshold density for rotation reversal in ECH power scan shots. The normalized effective collisionality ($\nu_{eff}$) at $\rho \sim 0.5$ at cut-off density and threshold density is also plotted in this figure.

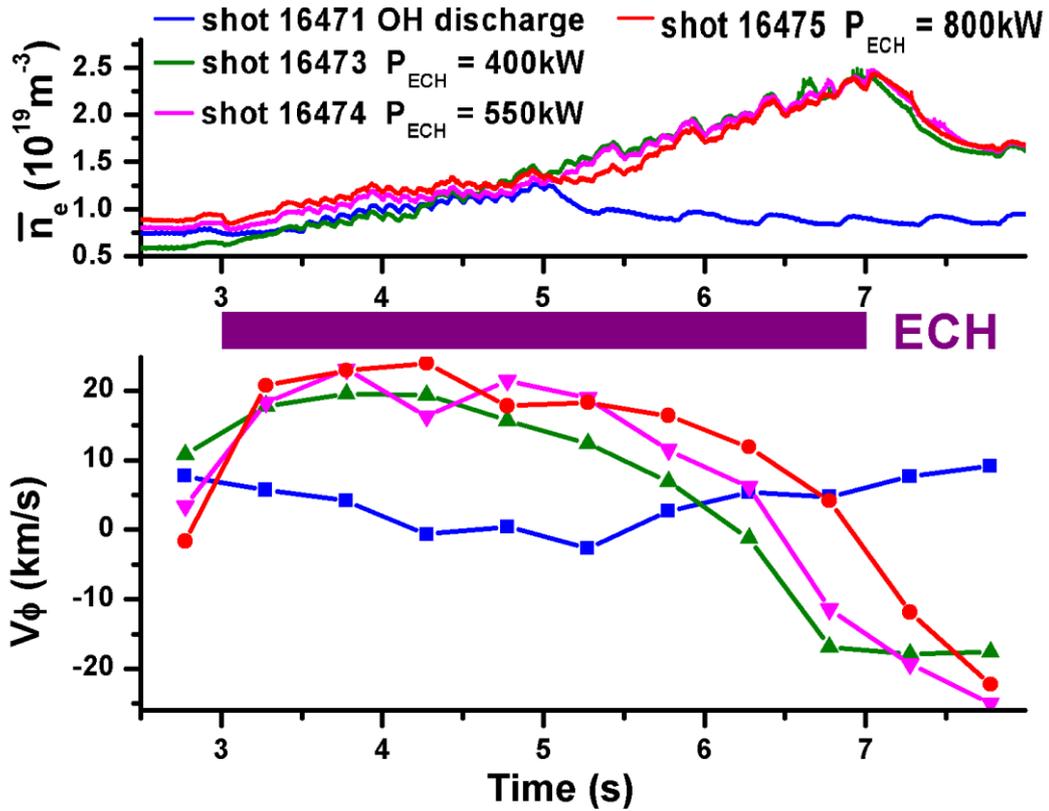

**Fig.5** Temporal evolutions of core $V\phi$ (measured by CES) and line averaged density for OH plasma and ECH plasmas with power scan

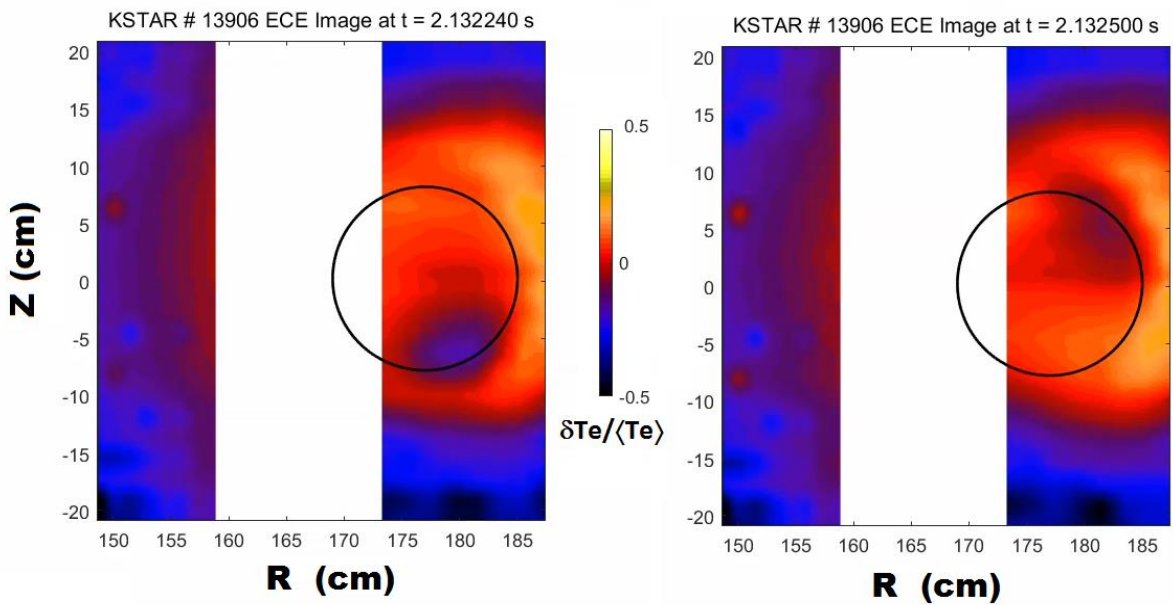

**Fig.6** 2D electron temperature fluctuation image from ECEI during NLT in ECH plasma. The time interval for the two image is 0.26ms. The pattern of temperature increasing area (red color) is very clear. The black circle represent the approximate location of q=1 surface. The blue spot at the q=1 surface is the m/n=1 island, which is in different location at different time due to the general rotated property of MHD island.

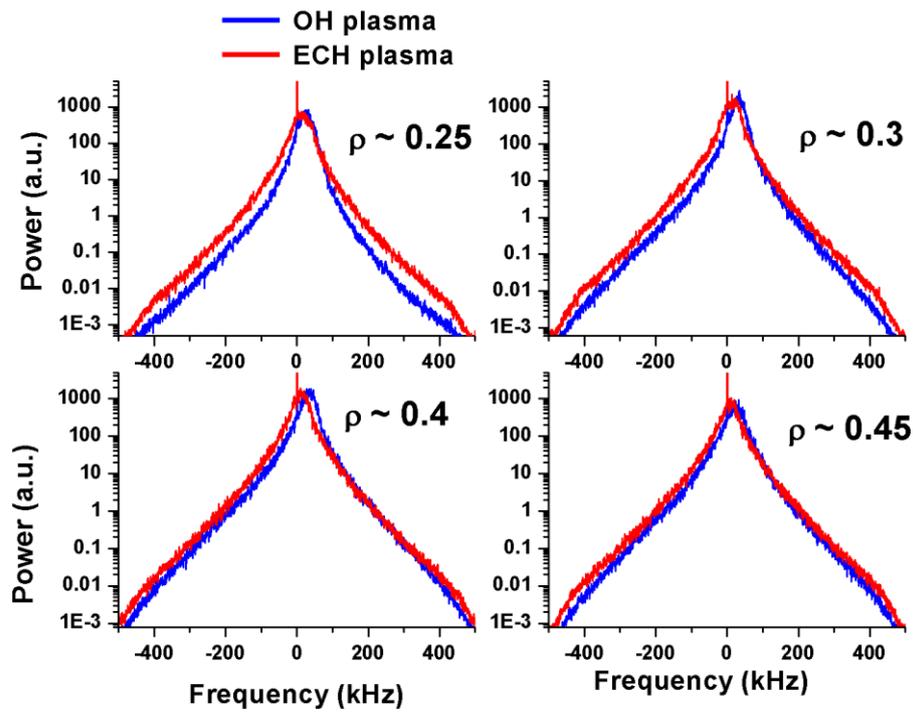

**Fig.7** Power spectra of density fluctuations in the ECH and OH plasmas. The time period in this figure is same as fig.3.

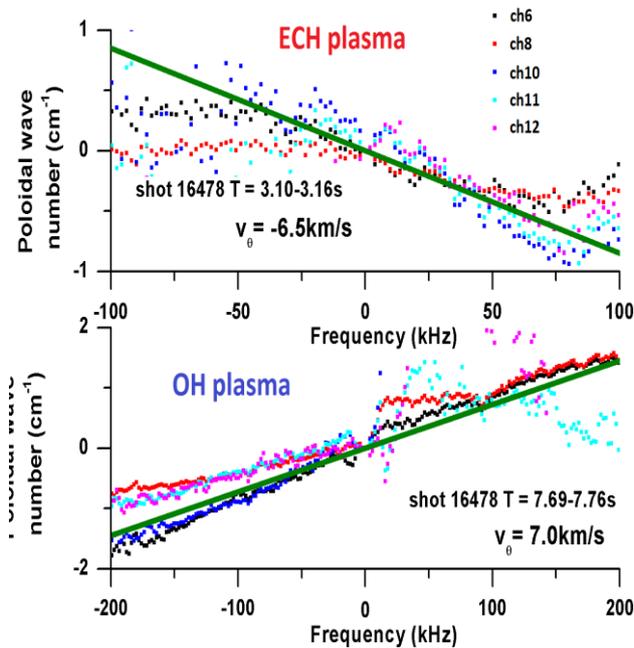

**Fig.8** The cross-phase spectra of MIR. Here, 5 dispersions for centre channels with respect to channel 9 are overlapped. The poloidal velocity is estimated from the inverse of the slope of a fitted line (green color) on the dispersion relation. The measurement position is about $\rho \sim 0.15$ for ECH plasma and $\rho \sim 0.35$ for OH plasma.

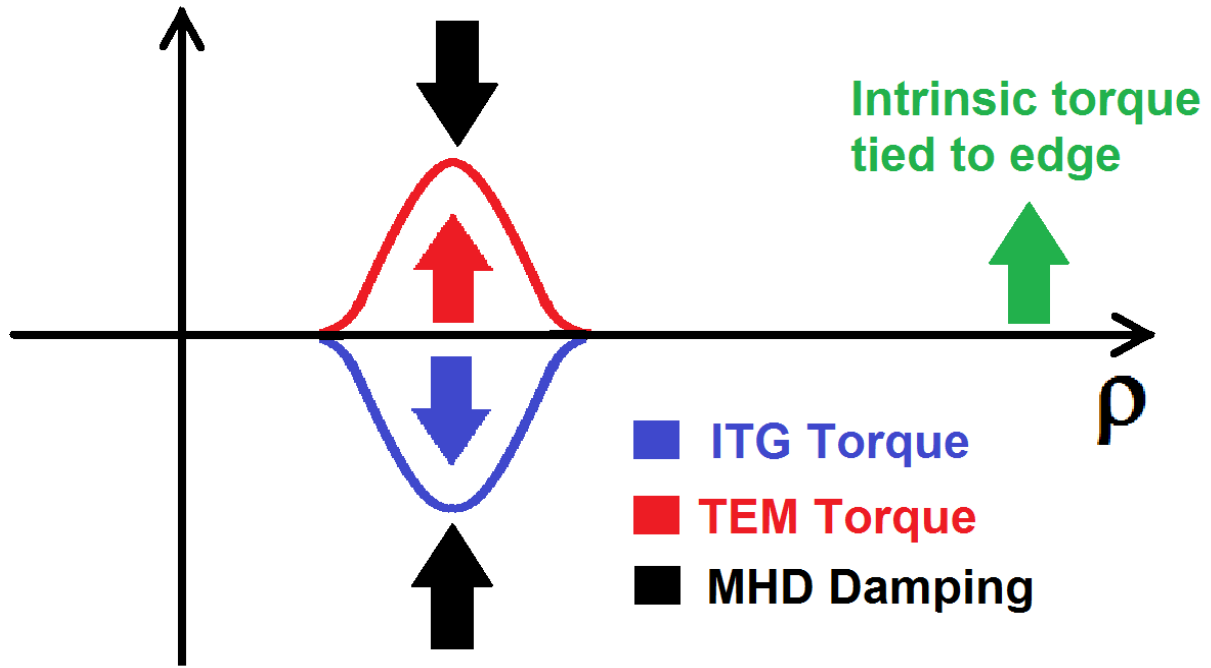

**Fig.9** Cartoons shows the torque and damping source for intrinsic rotation in KSTAR

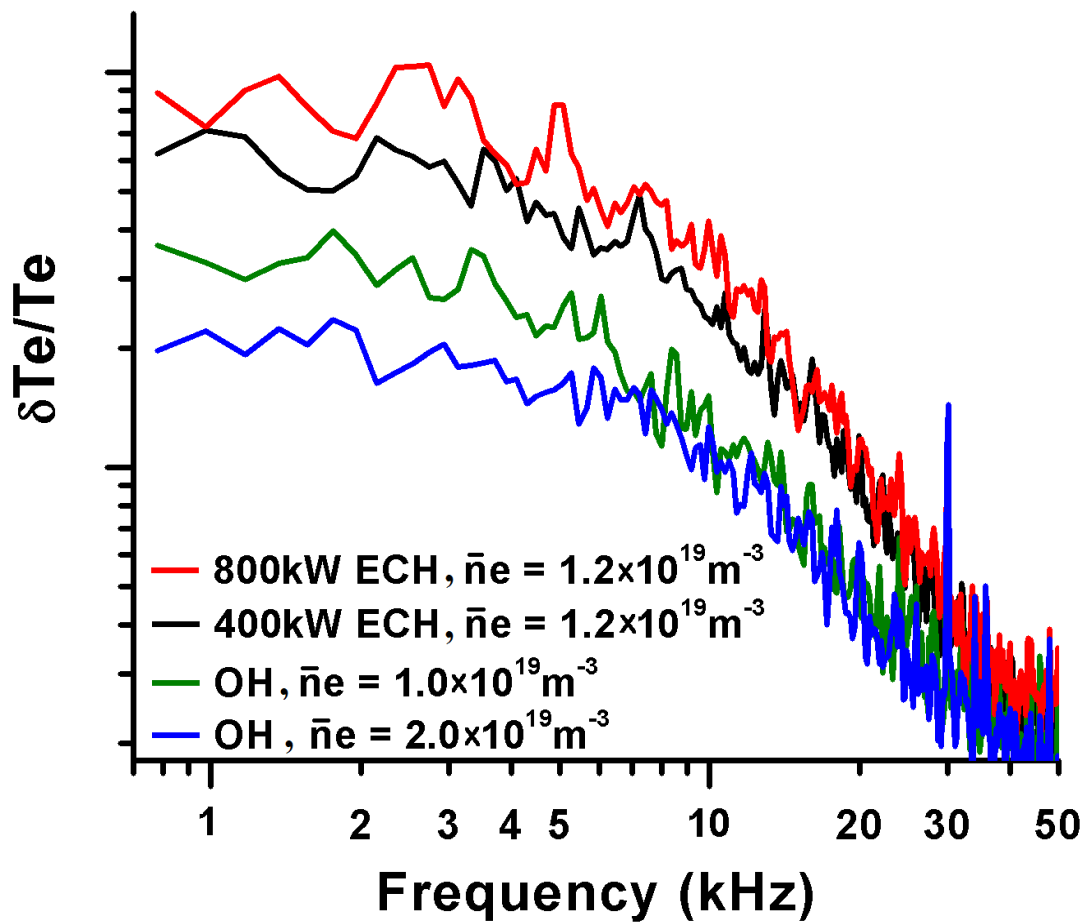

**Fig.10** Fluctuation power spectrum of Te at $\rho\sim 0.5$. The data of 400kW ECH is obtained from shot 16473. The other data are taken from shot 16475.

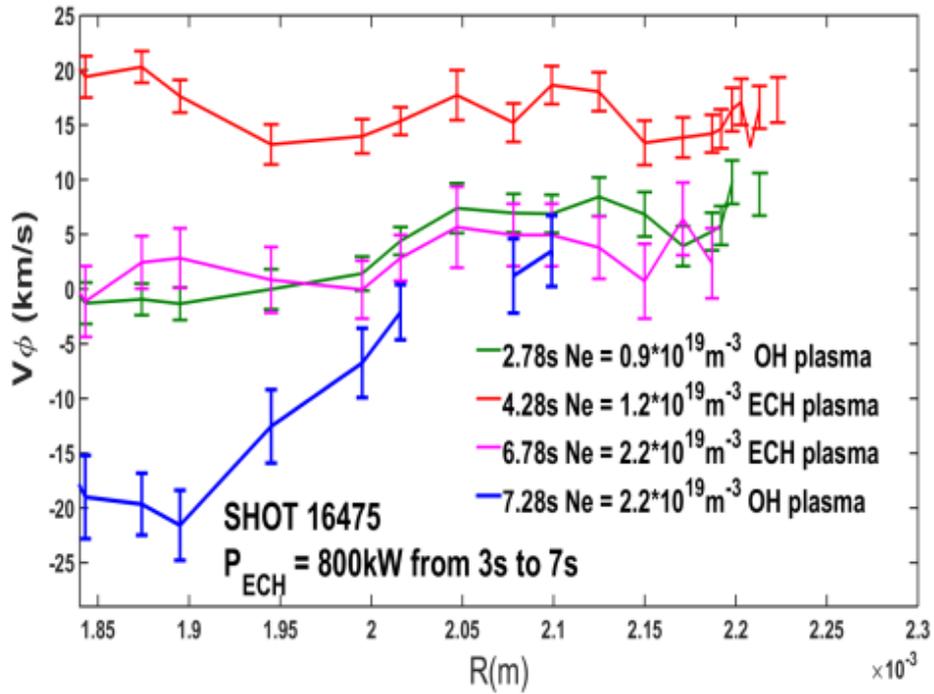

**Fig.11** Rotation profiles in shot 16475.

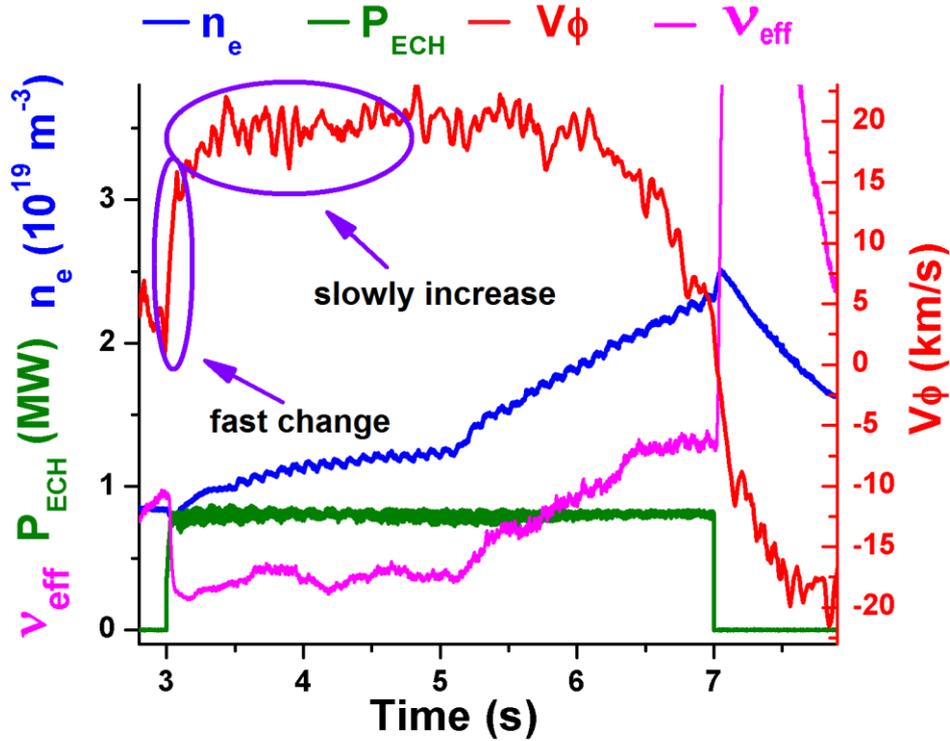

**Fig.12** Temporal evolution of core $V_\phi$, line averaged density, power of ECH, and $\nu_{eff}$ at $\rho\sim 0.5$ in shot 16478

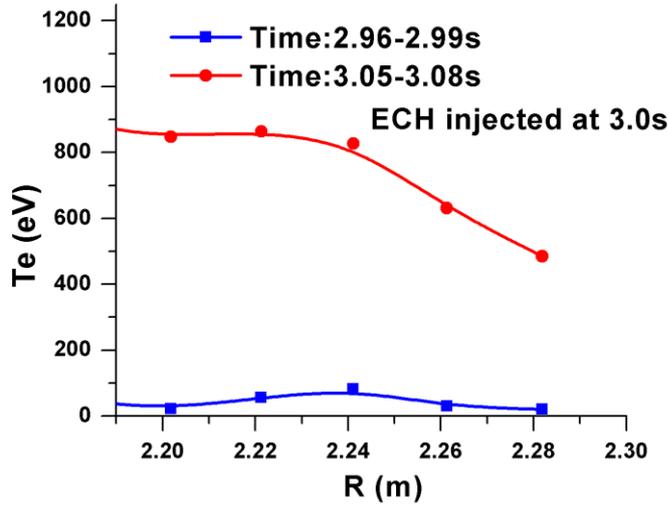

**Fig.13** Profiles of electron temperature at edge region in shot 16478. ECH is injected at 3.0s. The last closed flux surface is at about R=2.24m estimated by EFIT.

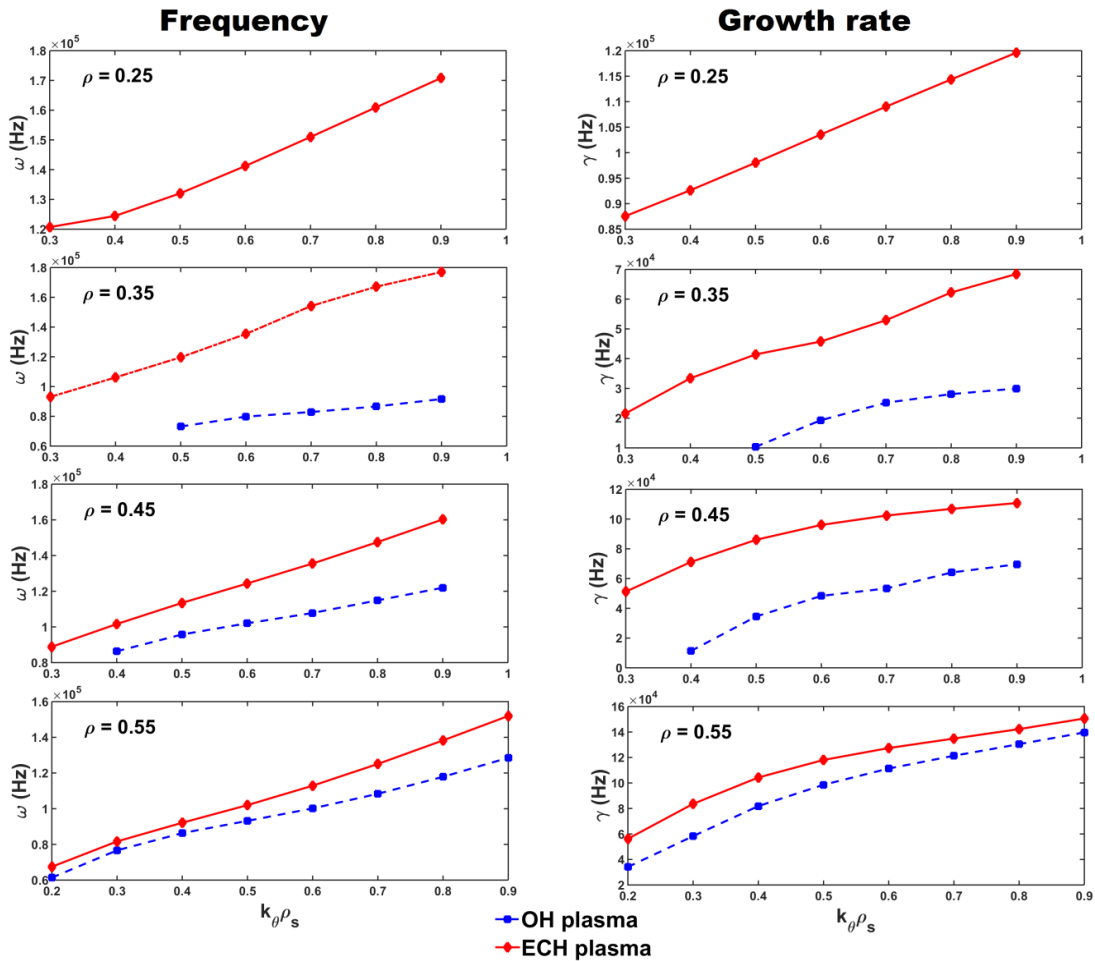

**Fig.14** The mode frequency and growth rate for OH plasma and ECH plasma. The experimental parameters for simulation are come from fig.3. Positive mode frequency means the modes in the electron diamagnetic direction; negative mode frequency means the modes in the ion diamagnetic direction.